\documentclass[12pt]{article}
\usepackage{times}
\usepackage{geometry}
\usepackage[numbers]{natbib}
\usepackage[utf8]{inputenc}

\usepackage{graphicx}

\usepackage{enumitem,amssymb}
\usepackage{ragged2e}
\usepackage{cancel}
\usepackage{comment}

\newlist{thematic}{itemize}{8}
\setlist[thematic]{label=$\square$}
\usepackage{pifont}
\newcommand{\cmark}{\ding{51}}%
\newcommand{\done}{\rlap{$\square$}{\raisebox{2pt}{\large\hspace{1pt}\cmark}}%
\hspace{-2.5pt}}

\usepackage{aas_macros}
\usepackage{xcolor}
\usepackage{soul}

\newcommand{\kb}[1]{\textbf{\color{teal} #1}} 

\begin{document}
\raggedright
\Large
Astro2020 Science White Paper\linebreak

\huge
Stellar multiplicity: an interdisciplinary nexus\linebreak
\normalsize

\noindent \textsl{Thematic Areas:} \hspace*{60pt} $\done$ Planetary Systems \hspace*{10pt} $\done$ Star and Planet Formation \hspace*{20pt}\linebreak
$\done$ Formation and Evolution of Compact Objects \hspace*{26pt} $\done$ Cosmology and Fundamental Physics \linebreak
  $\done$  Stars and Stellar Evolution \hspace*{1pt} $\done$ Resolved Stellar Populations and their Environments \hspace*{40pt} \linebreak
  $\done$    Galaxy Evolution   \hspace*{45pt} $\done$             Multi-Messenger Astronomy and Astrophysics \hspace*{65pt} \linebreak
  
\textsl{Principal Authors:}

Name: \textbf{Katelyn Breivik}
 \linebreak						
Institution:  Canadian Institute for Theoretical Astrophysics
 \linebreak
Email: kbreivik@cita.utoronto.ca
 \linebreak
Name: \textbf{Adrian M. Price-Whelan}
 \linebreak						
Institution: Princeton University
 \linebreak
Email: adrn@astro.princeton.edu
 \linebreak

\textsl{Co-authors:} 
\textbf{Daniel~J.~D'Orazio} (Harvard University),
\textbf{David~W.~Hogg} (New York University; MPIA; Flatiron Institute),
\textbf{L.~Clifton~Johnson} (Northwestern University),
\textbf{Maxwell~Moe} (University of Arizona),
\textbf{Timothy~D.~Morton} (University of Florida; Flatiron Institute),
\textbf{Jamie~Tayar} (University of Hawai'i; Hubble Fellow).
\linebreak

\textsl{Abstract:}
Our uncertainties about binary star systems (and triples and so on) limit our capabilities in literally every single one of the \textsl{Thematic Areas} identified for Astro2020.
We need to understand the population statistics of stellar multiplicity and their variations with stellar type, chemistry, and dynamical environment:
Correct interpretation of any exoplanet experiment depends on proper treatment of resolved and unresolved binaries;
stellar multiplicity is a direct outcome of star and companion formation;
the most precise constraints on stellar structure come from well-characterized binary systems;
stellar populations heavily rely on stellar and binary evolution modeling;
high-redshift galaxy radiation and reionization is controlled by binary-dependent stellar physics; 
compact objects are the outcomes of binary evolution;
the interpretation of multi-messenger astronomy from gravitational waves, light, and neutrinos relies on understanding the products of binary star evolution;
near-Universe constraints on the Hubble constant with Type Ia supernovae and gravitational-wave mergers are subject to systematics related to their binary star progenitors;
local measures of dark-matter substructure masses are distorted by binary populations.
In order to realize the scientific goals in each of these themes over the next decade, we therefore need to understand how binary stars and stellar multiplets are formed and distributed in the space of masses, composition, age, and orbital properties, and how the distribution evolves with time. This white paper emphasizes the interdisciplinary importance of binary-star science and advocates that coordinated investment from all astrophysical communities will benefit almost all branches of astrophysics.
\thispagestyle{empty}

\pagebreak
\setcounter{page}{1}

\RaggedRight
\setlength\parindent{1.4em}

\noindent{\textbf{{\large{Recommendation}}}}
Binary stars, and their evolution across the H-R Diagram, are critical to nearly all aspects of Astronomy. There are no technological or observational barriers to a near-complete observational determination of the full distribution of stellar mutiplicity as a function of stellar type, composition, environment, and orbital architecture. Given the interdisciplinary value of such results, the astrophysical community could and should join together to solve this problem. 

In practice, this will require comprehensive observational and theoretical analysis with photometric, spectroscopic, and astrometric surveys in concert.
Many existing or planned projects (e.g., LSST, SDSS-V, Gaia, ALMA, LIGO, LISA) have the capacity to deliver a subset of the required data (modulo survey execution decisions), with the largest missing component being time-resolved, multiplexed, infrared spectroscopy on a large aperture telescope.
To emphasize the concreteness of what we are recommending, we estimate that a coordinated, cooperative effort on stellar multiplicity (given existing or planned surveys) would require a total investment comparable to a 5-year, ground-based spectroscopic survey on an existing 4m-class telescope---i.e. far less than a major mission or facility and \emph{a shared cost across all astrophysical communities}. 
Such observations, combined with theoretical developments, would provide the information necessary to bring down scientific barriers in every one of the Astro2020 \textsl{Thematic Areas}.
\linebreak

\noindent{\textbf{{\large{Planetary Systems}}}}
The last decade of exoplanetary science has demonstrated that most planetary systems are unlike our own, both in terms of their orbital architecture and in the masses and radii of their typical constituents.
A continuing goal for this field is then to characterize the demographics of planetary systems: The occurrence rates of planets as a function of their mass, orbit, and host star parameters.
However, \textbf{measurements of individual planetary properties and planetary population constraints from exoplanet surveys are and will continue to be biased by presently unknown multiple star contamination.}

Of the thousands of exoplanetary systems now known, the vast majority have been discovered from planetary transits.
Planetary transits provide key information about planet radii, which are used to study the composition and structure of exoplanets.
Unresolved stellar multiplicity can affect planetary transit surveys in two key ways.
First, if the measured light during a transit comes from more than one star (e.g., a blended binary star system), the depth of this transit will be diluted, leading to an underestimate of the planet's size \citep[the magnitude of which varies from a few percent to a large factor; e.g.,][]{Torres:2004}.
Second, the apparent fractional dimming attributed to a transiting exoplanet could instead be caused by a background or grazing eclipsing binary system \citep[e.g.,][]{Poleski:2010}.
For individual systems, the presence of a stellar companion can be tested with detailed follow-up observations or precisely-measured light curves \citep[e.g.,][]{Hirsch:2017}.
However, for large transit surveys such as the Kepler, TESS, or planned PLATO missions, the vast majority of exoplanet candidates are not followed up.
Inferring the exoplanet demographics from large surveys therefore requires an accurate model of the population of binary stars.

Population synthesis models of these and more subtle effects for large transit surveys have shown that existing occurrence rates of large planets ($\gtrsim 2\,\rm{R}_{\oplus}$) may only be incorrect by a few percent, but the occurrence rate of small planets could be overestimated by as much as $\sim$$50\%$, depending on their intrinsic size distribution \citep{Bouma2018}.
However, the magnitude of this bias depends critically on the details of the binary and multiple star population characteristics, which are still poorly characterized beyond the solar neighborhood.

Future exoplanet population studies will continue to be inescapably affected by unresolved binary stars. \textbf{To infer the true distribution of planetary characteristics and connect these inferences with theories of planet formation and evolution we must simultaneously model planets and stellar multiplicity.} 
\linebreak

\noindent{\textbf{\large{Star and Planet Formation}}}
Most stars are initially born in multiple star systems; the subsequent
rapid evolution of which delivers the stellar multiplicity we observe in the field.
\textbf{Measuring the properties of binary stars as a function of age (and especially at very young ages), environment, and orbital separation will therefore provide the critical data required to develop an empirically-solid understanding of star formation.}

For young stars, near-infrared imaging \citep{Duchene2007, Connelley2008} and radio interferometry \citep{Tobin2016} of select, highly embedded protostars reveal a significant wide binary fraction ($a > 500\,\rm{au}$) that is substantially larger than that observed in the field. These observations demonstrate that turbulent fragmentation of molecular cores is a highly efficient process and that subsequent dynamical processing quickly disrupts wide binaries on cluster-crossing timescales.  Adaptive-optics imaging of a handful of T Tauri stars reveals a similar binary excess across intermediate separations $a\sim10-100\,\rm{au}$ \citep{Ghez1993, Kraus2011}, which poses a significant challenge to models of binary formation and dynamical evolution \citep{Duchene2018}. Meanwhile, spectroscopic monitoring of T Tauri stars shows the close binary fraction ($a<10\,\rm{au}$) of late-type stars is consistent with the field population \citep{Mathieu1994, Melo2003}, suggesting fragmentation and migration within the primordial disk occurs during the highly embedded phase \citep{Moe2018a}. {\textbf{ALMA, improved adaptive optics imaging, and wide-field spectroscopic surveys are needed to shed further light on how stellar companions fragment within cores and disks and subsequently accrete and migrate during their first few Myr.}}
 
Once stars have reached their zero-age main sequence, it is observed that the binary fraction increases with stellar mass \citep{Raghavan2010, Sana2012, Duchene2013, Moe2017}\kb{.} 
However, other parameters such as metallicity and environmental density also play an important role. For example, the close binary fraction of solar-type stars varies by a factor of $\sim4$ across $-1.0<\rm{[Fe/H]}<0.5$ while the wide binary fraction appears to be metallicity invariant \citep{Badenes2018, Moe2018, El-Badry2019}. Time-domain surveys such as LSST and Gaia will help fill in the parameter space by discovering millions of eclipsing, spectroscopic, astrometric, and common-proper-motion binaries across a broad range of environments \citep{Prsa2011, Eyer2015}. \textbf{Follow-up spectroscopy will be required to fully characterize their properties}.
 
Binary dynamics ($a<50\,\rm{au}$) may suppress the formation of planets \citep[e.g.,][]{Wang2014, Ngo2016}. 
Binaries therefore bias or affect planet formation rates and the inferred trends with respect to host mass and metallicity \citep{Wang2015}.
\textbf{The complex inter-relationships between binarity, stellar composition, and the formation of hierarchical multiplicity architectures, all motivate careful causal statistical work, and subtle observational and theoretical interpretation.}
\linebreak

\noindent{\textbf{\large{\emph{Single} Stars and Stellar Evolution}}}
Despite nearly a century of study, understanding \emph{single star} evolution across the H-R Diagram is still a challenge. 
Observations of both low and high mass stars show deviations from predicted masses and ages when comparing to theoretical models \citep[e.g.,][]{Mann2015, David2019}. 
Even with exquisite precision from astroseismology \citep[e.g.,][]{Tayar2017} or cluster fitting \citep[e.g.,][]{Choi2018}, the predicted temperatures of stars on the giant branch show large deviations from theory. 
These tensions are further compounded when considering mass loss or mixing that occurs during post-main sequence evolution.
\textbf{Binary star systems (especially non-interacting systems) provide empirical benchmarks for cross-calibrating stellar models across stellar type and evolutionary phase.}

Single star evolution has long been explored with observations of coeval stellar populations in open clusters. However, it has become clear that the limited dynamic range in the ages and chemical abundances of open clusters leaves significant gaps in our tests of stellar evolution models.
Stellar multiples, which necessarily have the same age and composition, serve as the smallest open clusters, and as an ensemble, are able to constrain stars of all types. 
Wide binary stars (and higher order multiples) with precise and accurate measurements of the masses, abundances, temperatures, gravities, and radii can indicate that their stellar components indeed have the same age and composition \citep[\rm{e.g.}, ][]{Oh2018,Andrews2018}. 
This connection can be used as a tool where deviations between theoretically predicted stellar properties and observed properties of stars in binaries indicate the presence of physics lacking in stellar models, and patterns of deviations as a function of stellar parameters can inform the type of missing physics. 
Applying this technique to detached binaries across the H-R Diagaram provides an interlaced structure on which a thorough understanding of single star evolution can be built. \textbf{A comprehensive description of single star evolution therefore requires high-resolution spectroscopic observations of binary stars across large swaths of separations, masses, compositions and evolutionary phases}. 	 
\linebreak

\noindent{\textbf{\large{Resolved and Unresolved Stellar Populations}}}
Whether interpreting the color-magnitude diagrams of star clusters or deriving star formation history constraints for nearby galaxies, the field of resolved stellar populations relies heavily on stellar evolution models to interpret observations.
\textbf{ The use of single-star stellar evolution models and isochrone sets, excluding binary stars and their effects, is one of the most significant sources of unmodeled uncertainty and systematic error in the study of stellar populations}.

Omitting binary star effects in stellar population analysis is especially problematic when studying massive stars, since the majority of these stars are found in binary systems.  More than $70\%$ of O-type stars ($\rm{M} > 15 \rm{M}_{\odot}$) exchange mass with a companion during their lifetime \citep{Sana2012}. Excluding the effects of  accretion and mergers causes populations of rejuvenated young massive stars within star clusters to appear to be up to 2$\times$ more massive and 10$\times$ younger than the mass and age limits assumed from single star models. This leads to biases in mass function slope measurements, initial mass function constraints, and age determinations for these clusters \citep{Schneider2014, Schneider2015}.

The effects of binary stellar evolution similarly appear in composite, unresolved stellar populations, affecting constraints on galaxy formation and evolution. Binary-enabled pathways give rise to harder UV spectra as massive stripped and rejuvenated stars contribute additional Lyman and UV continuum emission \citep[e.g., ][]{Wilkins2016, Zackrisson2017}.  Also, the range of stellar ages over which UV emission is emitted significantly increases due to binary evolution, affecting age estimates and star formation rate calibrations based on UV continuum or nebular line emission \citep[e.g., ][]{Stanway2016}. Beyond massive stars, binary mass transfer interactions are responsible for blue straggler and extreme horizontal branch stars, which contribute UV and blue emission that would otherwise be unexplained in an old stellar population \citep[e.g., UV-upturn in elliptical galaxies; ][]{Han2007}.  

\textbf{Continued development and calibration of stellar evolution and population synthesis models with stellar multiplicity \citep[e.g.,][]{Eldridge2017} are required for interpreting observations of resolved and unresolved stellar populations across cosmic time.}
\linebreak

\noindent{\textbf{\large{Formation and Evolution of Compact Objects}}}
Stellar-mass compact objects are notoriously difficult to discover and characterize. Given the relatively small number of known systems (as compared to stellar populations), models for the formation and evolution of compact object populations remain largely unconstrained.
Binary star surveys help this effort in two ways: Finding stars with unseen companions provides a way to identify new compact objects, and understanding binary star interactions will help connect compact objects to their progenitor stars.
\textbf{Understanding the relationship between compact object progenitors and their binary companions is key to uncovering both the physical processes that shape binary star evolution as well as the rates and properties of compact object populations}.

Constraining models of the formation and evolution of compact objects will require finding large samples of binary systems with compact object members over a range of masses and separations.
Such samples would enable addressing major open questions like the nature of the underlying supernova engine \citep[e.g.,][]{Fryer2012,Pejcha2015,sukhbold2016}, the source and strength of natal kicks \citep[e.g.,][]{Kalogera1996,Hobbs2005,Bray2016}, and the mass distribution of black holes and neutron stars (which has uncertainties both at the low, $2-5\rm{M}_{\odot}$, and high, $40-60\rm{M}_{\odot}$, masses).
\textbf{Given the significant uncertainty in our current understanding of these processes, the identification of compact object members in binary star systems and the development of improved theoretical models for their evolution must be made a priority for the next decade.}
\linebreak 

\noindent{\textbf{\large{{Multi-Messenger Astronomy and Astrophysics}}}}
\textbf{The largest uncertainties in multi-messenger observations of compact binary populations lie in their astrophysical interpretation and are a direct product of deficiencies in our understanding of the evolution of their binary star progenitors}. The impact of common envelope evolution, the strength of compact object natal kicks, and the effects of dynamical encounters, among other processes, are not fully understood and are each responsible for \emph{orders of magnitudes} in uncertainty in the rates and characteristics of compact object populations. These uncertainties are further compounded through complicated interdependencies of different processes as well as uncertainties in the initial distributions of binary star masses, separations, eccentricities, compositions, and birth times. 

The recently discovered binary neutron star (NS) merger \citep{Abbott2017}, a flagship multi-messenger source of gravitational waves, neutrinos, and light across the electromagnetic (EM) spectrum, is a useful illustration in this context. This discovery delivered several incredible results, from confirming NS mergers as an origin of r-process elements \citep{Abbott2017_rprocess} to constraining the NS equation of state \citep{Abbott2018_NSeq}. However, explicit tuning to models for cosmological star formation history, initial binary distributions, and binary evolution models is required to reproduce a population with binary NS and binary black hole (BH) merger rates consistent with LIGO's observed rates \citep[e.g.,][]{Chruslinska2018, Eldridge2019}. Similar problems are present for the population of Galactic white dwarf (WD) binaries, expected to be observed by LISA and electromagnetic observatories \citep[e.g.,][]{Korol2017, Breivik2018}: Standard binary evolution models over-predict their abundance by an order of magnitude when compared to the observed space density of interacting white dwarfs \citep{Roelofs2007}.

As multi-messenger observations are made over the next decades, the rates and properties of binary populations containing WDs, NSs, and BHs will come into focus. 
\textbf{To fully realize the potential of multi-messenger astrophysics, the interpretation of these observations will require connecting the formation and evolution of binary stars to the compact object populations they produce.}
\linebreak

\noindent{\textbf{\large{Cosmology}}}
Observations of the Hubble constant (H$_{\rm{0}}$) show a tension between measurements from local ($z\lesssim2$) and cosmological distances, suggesting the need for new redshift and distance calibrations. In the nearby Universe, type Ia Supernovae (SNe Ia) are the most widely used calibrators. There are several proposed progenitor classes to SNe Ia, but most involve thermonuclear detonations of WDs in binaries that can be divided into two classes: single and double degenerate \cite{Maoz2014}. As catalogs of SNe grow thanks to ongoing and planned time-domain photometric surveys, \textbf{systematics at the $\sim2\%$ level from undetermined SNe Ia progenitors, and their evolution with metallicity and age, will influence precision cosmological measurements} \citep[\rm{e.g.}, ][]{Howell2011}. 

At redshifts $z<0.5$, gravitational-wave and multi-messenger observations of binary BH and binary NS mergers can independently constrain H$_{\rm{0}}$ to $\sim1-5\%$-level precision \cite{Schutz1986, Abbott2017_h0, Fishbach2019}. Cases where both light and gravitational waves are observed allow independent distance and redshift observations that can be used to calibrate population measurements with gravitational waves alone \citep{Chen2018}. Binary BH mergers at high masses near the proposed pair instability SN gap may also be used to calibrate H$_{\rm{0}}$ measurements with independent distance and redshift measurements at higher redshifts than binary NS mergers. In this case, masses that fall in the gap are a direct measure of the redshift effects on the BHs \citep{Farr2019}. \textbf{Precise H$_{\rm{0}}$ measurements with gravitational waves alone rely on confirmation of the pair instability gap and the systematic uncertainties that arise from the evolution and redshift dependent formation rates of very high mass ($\gtrsim$\,$100\rm{M}_{\odot}$) binary stars}.
\linebreak

\noindent\textbf{\large{Dark matter}}
Dwarf galaxies around the Milky Way and M31 provide critical information about the small-scale (mass $\lesssim 10^{10}~\rm{M}_{\odot}$) properties of dark matter physics \citep[e.g.,][]{Bullock:2017}. 
Low-mass galaxies provide unique laboratories to study mass-to-light ratios and mass profiles in dark matter environments that are expected to have been less affected by baryonic physics and feedback relative to Milky Way-mass galaxies. 
Given the typical distances ($\sim 10$--$1000\,\rm{kpc}$) to these dwarf galaxies, most of what is known about their dark matter distributions has therefore been inferred from line-of-sight velocity dispersions and dispersion profiles of stars in these dwarf galaxies \citep[e.g.,][]{Walker:2009}. 
Given the difficulty in obtaining these data, the inferred (kinematic) mass measurements of the lowest-mass galaxies known typically hinge on single-epoch or sparsely time-domain velocity data for small samples of stars; 
The interpretation of these velocity measurements can be significantly affected by the presence of unknown close binary star systems.
\textbf{Inferring robust dynamical (dark matter) masses for local group satellites requires marginalizing over the presently unknown stellar multiplicity statistics of dwarf galaxy environments.}

The presence of unknown close binary star systems in a sample of stars with single-epoch or sparse line-of-sight velocity data will generically increase the observed velocity dispersion of the system \citep[e.g.,][]{Olszewski:1996}.
This is particularly relevant for dwarf galaxies where the velocity dispersions are expected to be $\lesssim 5~\,\rm{km/s}$. However, it is currently not understood how binary star population characteristics depend on (chemical or dynamical) environment. For Milky Way stars, it has recently been shown that the binary fraction increases with decreasing metallicity \citep{Moe2018}, but it is not clear how generic this is for stellar populations with different star formation dynamics and histories \citep[i.e. dwarf galaxies;][]{Minor2013}. In fact, recent work has shown that even in similar chemical environments (i.e. amongst dwarf satellites of the Milky Way), the binary star population statistics may vary significantly \citep[e.g.,][]{Martinez2011, Spencer2018}.

\textbf{The unknown binary star population statistics in dwarf galaxies throughout the local group are therefore a source of significant uncertainty for constraining the small-scale or low-mass properties of dark matter.}
\linebreak

\noindent If you made it this far, and want closure: See the \textsl{Recommendation} at the start.

\pagebreak
\bibliographystyle{plainnat}

\end{document}